\documentstyle[10pt,aps,prd,float,epsf]{revtex}
\begin{document}
\newcommand{\om}[1]{\omega_{#1}}
\title{The Renormalized Thermal Mass with Non-Zero Charge Density}
\author{H.~F.~Jones\footnote{e-mail h.f.jones@ic.ac.uk} and Philip~Parkin\footnote{e-mail p.parkin@ic.ac.uk}}
\address{Physics Department,
    Imperial College, London, SW7 2BZ, UK}
\maketitle
\begin{abstract}
The linear $\delta$ expansion is used to obtain corrections up to ${\mathcal O}
(\delta^2)$ to the self-energy for a complex scalar field theory with a $\lambda
(\varphi^{\star}\varphi)^2$  interaction at high temperature and non-zero
charge density. The calculation is done in the imaginary-time formalism via the
Hamiltonian form of the path integral. Nonperturbative results are generated
by a systematic order by order variational procedure and the dependence
of the critical temperature on the chemical potential $\mu$ is obtained.
\smallskip

\end{abstract}

\section{Introduction}
It has been known for some time that conventional perturbation theory is
inadequate for describing high-temperature field theory \cite{lin79}, with
the perturbation expansion breaking down at some order in the coupling
constant in the parameter regime where the temperature is very much greater
than the bare mass.
There have been several attempts to circumvent this problem (for example
Refs.~\cite{par92,ban91} use resummed perturbative expansions and
systematically include all relevant diagrams). A
temperature-dependent renormalization group approach has been used
in Ref.~\cite{fun87}.  The method we will employ in this paper is the
linear $\delta$ expansion (see, for example \cite{buc93}), which
contains elements of both of the above methods. It involves a mass shift
which is determined order-by-order in the expansion by a
non-perturbative criterion, but prior to the (crucial) optimization
stage it merely uses low-order perturbation theory with modified propagators
and vertices.

Pinto and Ramos \cite{pin99} have successfully applied the method
to finite-temperature symmetry breaking in a scalar field theory
with a $\varphi^4$ interaction up to second order in $\delta$,
which in particular involves evaluating the non-trivial ``setting-sun"
diagram. In the present paper we extend this approach
to tackle the problem of a non-zero chemical potential;
in so doing we consider a complex scalar
field rather than a real scalar field. For $\mu=0$, the
effect is simply to alter the symmetry factors of the one- and two-loop
diagrams, but for $\mu\ne 0$ there is a shift in the energy component of the
Matsubara propagator, which makes the diagrams considerably more
difficult to evaluate but presents no problem of principle.

To our knowledge there have been only a few attempts at this
problem. Benson et al. \cite{ben91} performed a one-loop
calculation, while Funakubo and Sakamoto \cite{fun86} used
a renormalization group approach in the large-$N$ limit of
an ${\rm O}(N)$ theory (in which the setting-sun diagram does
not appear). In the standard lattice Monte-Carlo approach
it is impossible to incorporate a non-zero chemical potential, because the
Euclidean action then becomes complex and cannot be used as a real, positive
statistical weight.

We outline the linear $\delta$ expansion technique below and explain how
it can generate nonperturbative results, while in Sec.\ II we present
briefly the construction of the theory in the imaginary time formalism and
include the chemical potential associated with a conserved charge.  We then
go on to calculate all diagrams contributing to the self-energy, up to
second order in $\delta$. Sec.\ III includes results for the critical
temperature and its behaviour as one allows $\mu$ to take non-zero values.
The appendix contains a detailed calculation of the two-loop setting sun
diagram in the high temperature expansion.
\subsection{The Linear Delta Expansion}
The linear $\delta$ expansion (LDE) is a technique that allows the use of an
analytic approach to probe the nonperturbative sector of the field theory to
which it is applied. It has been employed with success in a wide variety of
areas \cite{buc93}, with convergence of the expansion rigorously demonstrated
in some simple zero- and one-dimensional models (see, for example,
\cite{arv95}). The method involves the introduction of an artificial expansion
parameter, $\delta$, and the modification of the action of the theory under
consideration via
\begin{equation}
    S \to S^{\delta}=(1-\delta)S_0(\{\eta_i\})+\delta S,
\end{equation}
where $\{\eta_i\}$ is some set of variational parameters and $S_0$
represents the action of some soluble theory. This trial action is
not determined by the method; however, an appropriate choice is usually
suggested by the form of the theory under consideration.
The $\delta$-modified action can then be used to evaluate some desired
physical quantity as a power series in $\delta$ (truncated at some
finite order, $N$), with $\delta$ then set equal to 1 at the end of the
calculation. We label this quantity $P_N$, noting that it will, in general,
have a residual dependence on the $\{\eta_i\}$. The variational element is
introduced by fixing these parameters according to some specified criterion.
We shall use the {\it principle of minimal sensitivity} (PMS) \cite{ste81}
whereby the $\{\eta_i\}$ are chosen at a stationary point of the quantity
$P_N$, namely:
\begin{equation}
    \left.\frac{\partial P_N}{\partial\eta_i}\right|_{\eta_i=
    \bar{\eta}_i}=0.
\end{equation}
It is this {\it order by order} fixing that allows for non-perturbative
behaviour to emerge (the optimized variational parameters becoming functions
of the order of truncation, and so being more correctly labelled by
$\{\eta_i(N)\}$), and can also provide for convergence of the expansion, two particularly
desirable features.
\section{The Charged Scalar Field At Finite Temperature}
In Minkowski space, the Lagrangian density of a massive, complex scalar field
$\varphi(x)$ with a ${\mathcal V}(\varphi^{\star}\varphi)$ interaction term is
given by
\begin{equation}
    {\mathcal L}=(\partial^{\mu}\varphi^{\star})(\partial_{\mu}\varphi)
    -m^2\varphi^{\star}\varphi-{\mathcal V}(\varphi^{\star}\varphi).
\end{equation}
This Lagrangian density has a well-known global $U(1)$ gauge symmetry which
leads to a conserved charge $\mathcal Q$  and an associated chemical potential
$\mu$, so that the grand partition function is
\begin{equation}
    Z(\beta,\mu)={\mathrm Tr}\:e^{-\beta(\hat{H}-\mu\hat{Q})}.
\end{equation}
Proceeding through the Hamiltonian form of the path integral for $Z$, in the
imaginary-time formalism of finite-temperature field theory~\cite{leb96}, we
arrive at
\begin{equation}
    Z(\beta,\mu)=\int{\mathcal D}(\varphi^{\star},\varphi)
    e^{-S_E(\beta,\mu)},
\end{equation}
where the Euclideanized action is (dropping the suffix for future notational
convenience)
\begin{equation}
    S(\beta,\mu)=S_F(\beta,\mu)+S_I(\beta,\mu)+S_C(\beta,\mu),
\end{equation}
with
\begin{eqnarray}
    S_F(\beta,\mu) &=& \int_T\!d^4\!x\,\varphi^{\star}\left
    [-\frac{\partial^2}{\partial\tau^2}+2\mu\frac{\partial}{\partial\tau}
    -\nabla^2+m^2-\mu^2\right]\varphi, \\
    S_I(\beta,\mu) &=& \int_T\!d^4\!x\,{\mathcal V}(\varphi^{\star}
    \varphi), \\
    S_C(\beta,\mu) &=& \int_T\!d^4\!x\,\varphi^{\star}\left
    [-A\left(\frac{\partial^2}{\partial\tau^2}-2\mu\frac{\partial}
    {\partial\tau}+\nabla^2+\mu^2\right)+Bm^2\right]\varphi+\int_T
    \!d^4\!x\,C{\mathcal V}(\varphi{\star}\varphi),
\end{eqnarray}
where $\tau=it$, $x=(\tau,{\mathbf x})$ and
$\int_T\!d^4\!x=\int_0^{\beta}\!d\tau\int\!d^3{\mathbf x}.$
Here $S_C(\beta,\mu)$ represents those counterterms required to render
the model finite.

A sensible choice for our delta-modified action, in the presence of a
$(\varphi^\star\varphi)^2$ self-interaction,
${\mathcal V}(\varphi^\star\varphi)=
(\lambda/4)(\varphi^\star\varphi)^2$,
would then be
\begin{equation}
    S^{\delta}(\beta,\mu)=S^{\delta}_F(\beta,\mu)+S^{\delta}_I(\beta,\mu)
    +S^{\delta}_C(\beta,\mu),
\end{equation}
with
\begin{eqnarray}
    S^{\delta}_F(\beta,\mu) &=& \int_T\!d^4\!x\,\varphi^{\star}
    \left[-\frac{\partial^2}{\partial\tau^2}+2\mu\frac{\partial}{\partial
    \tau}-\nabla^2+\Omega^2-\mu^2\right]\varphi, \\
    S^{\delta}_I(\beta,\mu) &=& \int_T\!d^4\!x\,\left[\frac{\delta
    \lambda}{4}(\varphi^{\star}\varphi)^2-\delta\eta^2\varphi^
    {\star}\varphi\right], \\
    \label{SCd}
    S_C^{\delta}(\beta,\mu) &=& \int_T\!d^4\!x\,\varphi^{\star}
    \left[-A^{\delta}\left(\frac{\partial^2}{\partial\tau^2}-2\mu\frac
    {\partial}{\partial\tau}+\nabla^2+\mu^2\right)+B^{\delta}\Omega^2
    \right]\varphi+\int_T\!d^4\!x\,\left[C^{\delta}\frac{\delta
    \lambda}{4}(\varphi^{\star}\varphi)^2-B^{\delta}\delta\eta^2
    \varphi^{\star}\varphi\right],
\end{eqnarray}
introducing the variational parameter $\Omega$, or equivalently
$\eta$, given by $\eta^2=\Omega^2-m^2$.
In subsequent sections we will evaluate the thermal mass up to
${\mathcal O}(\delta^2)$, using dimensional regularization to renormalize the
theory. It is an important point, raised in \cite{pin99}, that the
arbitrary variational parameter $\eta$ becomes a function of the bare
parameters (which is how non-perturbative behaviour arises in
the LDE), so we must eliminate any divergences before we apply the
PMS optimization procedure. The renormalization procedure we use is identical
to that of \cite{pin99} and is based on \cite{chi98}; we forgo any detailed
discussion of the cancelling of temperature-dependent divergences and any
systematic calculation of the coefficients of the counterterms $A^{\delta}$, $B^{\delta}$ and
$C^{\delta}$, the details being essentially identical to those discussed in
\cite{pin99}.

\subsection{Calculating the thermal mass}
We recall that the Feynman rules in frequency space in the presence of
an overall charge are as follows:
\begin{enumerate}
    \item To every line of the diagram assign a factor $\Delta_F(i\om{n},
    {\mathbf k})\equiv [-(i\omega_n+\mu)^2+\omega_k^2]^{-1}$, where $\omega_n=2\pi
    n/\beta$ and $\omega_k=\surd{({\mathbf k}^2+\Omega^2)}$,
    and an arrow in the direction of momentum flow
    \item Assign a factor $-\delta\lambda$ to each vertex
    \item Assign a factor $\delta\eta^2$ to each insertion
    \item Integrate over every internal line with the measure
    $T\sum_n\!\int\!{d^3{\mathbf k}}/{(2\pi)^3}$
    \item There must be conservation of charge at each vertex, i.~e. the
    number of arrows entering a vertex must be the same as the number
    leaving
\end{enumerate}
We are now in a position to estimate corrections to the thermal mass (which
we do up to ${\mathcal O}(\delta^2)$), defined by
\begin{equation}
   m_{T,\mu}^2=\Omega^2+\Pi(i\omega_n,{\mathbf p}),
\end{equation}
evaluated on-shell, i.e. $i\omega_n=\Omega-\mu,
{\mathbf p}={\mathbf 0}$, where $\Pi$ is the thermal self-energy.
This choice of four-momentum is discussed in the Appendix, where
the setting-sun diagram is calculated, the only momentum-dependent
contribution to $\Pi$ up to ${\mathcal O}(\delta^2)$.

\subsection{The thermal mass at first order}

\begin{figure}
%    \begin{center}
    \centerline{\epsfxsize=3cm
    \epsffile{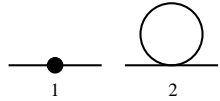}}
    \caption{Diagrams contributing at ${\mathcal O}(\delta)$}
%    \end{center}
\end{figure}

To lowest order, the relevant contributions are
\begin{eqnarray}
\label{Pib}
    \Pi_{T,\mu}^{\delta} & = & \Pi_1^{\delta}+\Pi_2^{\delta}
    \nonumber\\
    & = & -\delta\eta^2+\delta\lambda T\sum_n
    \int\!\frac{d^3{\mathbf k}}{(2\pi)^3}\Delta_F(i\omega_n,{\mathbf k}).
\end{eqnarray}
The frequency sum in (\ref{Pib}) can be treated in a particularly concise
and efficient manner through the mixed representation~\cite{pis88}
$\Delta_F(\tau,{\mathbf k})$ of $\Delta_F$:
\begin{eqnarray}
    \Delta_F(\tau,{\mathbf k}) & = & T\sum_ne^{-i\omega_n\tau}
    \Delta_F(i\omega_n,{\mathbf k}) \nonumber\\
    & = & \frac{1}{2\omega_k}\left[(1+n_k^-)
    e^{-(\omega_k-\mu)\tau}+n_k^+e^{(\omega_k+\mu)\tau}\right],
\end{eqnarray}
where
\begin{equation}
    n_k^{\pm}=\frac{1}{e^{\beta(\omega_k\pm\mu)}-1}
\end{equation}
is the Bose-Einstein distribution function in the presence of a non-zero
chemical potential.
The Matsubara propagator is recovered by Fourier transforming the mixed
propagator with respect to the $\tau$ variable:
\begin{equation}
\label{Del}
    \Delta_F(i\omega_n,{\mathbf k})=\int_0^{\beta}\!d\tau e^{i\omega_n\tau}
    \Delta_F(\tau,{\mathbf k}).
\end{equation}
The contribution of $\Pi_2^{\delta}$ in (\ref{Pib}) can then be calculated
trivially to give
\begin{eqnarray}
    \Pi_2^{\delta} & = &  \delta\lambda T\sum_n\int\!\frac{d^3
    {\mathbf k}}{(2\pi)^3}\int_0^{\beta}\!d\tau e^{i\omega_n\tau}\Delta_F
    (\tau,{\mathbf k}) \nonumber\\
    & = & \delta\lambda T\int\!\frac{d^3{\mathbf k}}{(2\pi)^3}\frac{1}
    {2\omega_k}(1+n_k^-+n_k^+).
\end{eqnarray}
Using dimensional regularization~\cite{col83} one finds that the ${\mathcal O}
(\delta)$ contribution to the thermal mass is
\begin{equation}
    m_{T,\mu}^2=\Omega^2-\mu^2-\delta\eta^2+\delta\frac{\lambda}
    {16\pi^2}\Omega^2\left[-\frac{1}{\epsilon}+\ln\left(\frac{\Omega^2}
    {4\pi M^2}\right)+\gamma-1\right]+\delta\frac{\lambda T^2}
    {\pi^2}h_3^e(y,r),
\end{equation}
where $M$ is a mass scale introduced by dimensional regularization and,
in the notation of~\cite{hab82},
\begin{eqnarray}
\label{hey}
    h_3^e(y,r) & = & {\textstyle\frac{1}{2}}[h_3(y,r)+h_3(y,-r)],\\
    h_3(y,r) & = & \frac{1}{\Gamma(3)}\int_0^{\infty}\!dx\frac{x^2}
    {\sqrt{x^2+y^2}}\left[\frac{1}{\exp(\sqrt{x^2+y^2}-ry)-1}\right],
\end{eqnarray}
where $x=\beta k$, $y=\beta\Omega$ and $r=\mu/\Omega$.  Using
the minimal subtraction scheme (MS) we can eliminate the divergent term
arising from the temperature-independent part of the self energy using the
${\mathcal O}(\delta)$ mass counterterm
\begin{equation}
    \Pi_{\mathrm{ct}}^{\delta}=\left(\delta\frac{\lambda}{16\pi^2
    \epsilon}\right)\Omega^2.
\end{equation}

We now analyze~(\ref{hey}) in the limit $y\to 0$ (the high temperature
expansion) keeping $r$ fixed~\cite{hab82}:
\begin{equation}
    h_3^e(y,r)=\frac{\pi^2}{12}-\frac{\pi y}{4}\sqrt{1-r^2}-\frac{y^2}{8}
    \left[\ln\left(\frac{y}{4\pi}\right)+\gamma-{\textstyle\frac{1}{2}}
    +r^2\right]+\cdots,
\end{equation}
leading to the following expression for the renormalized thermal mass at first
order:
\begin{equation}
    m^2_{T,\mu}=\Omega^2-\delta\eta^2+\delta\frac{\lambda T^2}
    {12}-\delta\frac{\lambda T\Omega}{4\pi}\sqrt{1-r^2}+\delta\frac
    {\lambda\Omega^2}{16\pi^2}\left[\ln\left(\frac{4\pi T^2}{M^2}
    \right)-\gamma-2r^2\right].
\end{equation}
In this and all subsequent calculations we neglect terms of ${\mathcal
O}(1/T)$.
%%%%%%%%%%%%%%%%%%%%%%%%%%%%%%%%%%%%%%%%%%%%%%%%%%%%%%%%%%%%%%%%%%%%%%
\subsection{The thermal mass at second order}

\begin{figure}
%    \begin{center}
    \centerline{\epsfxsize=8cm
    \epsffile{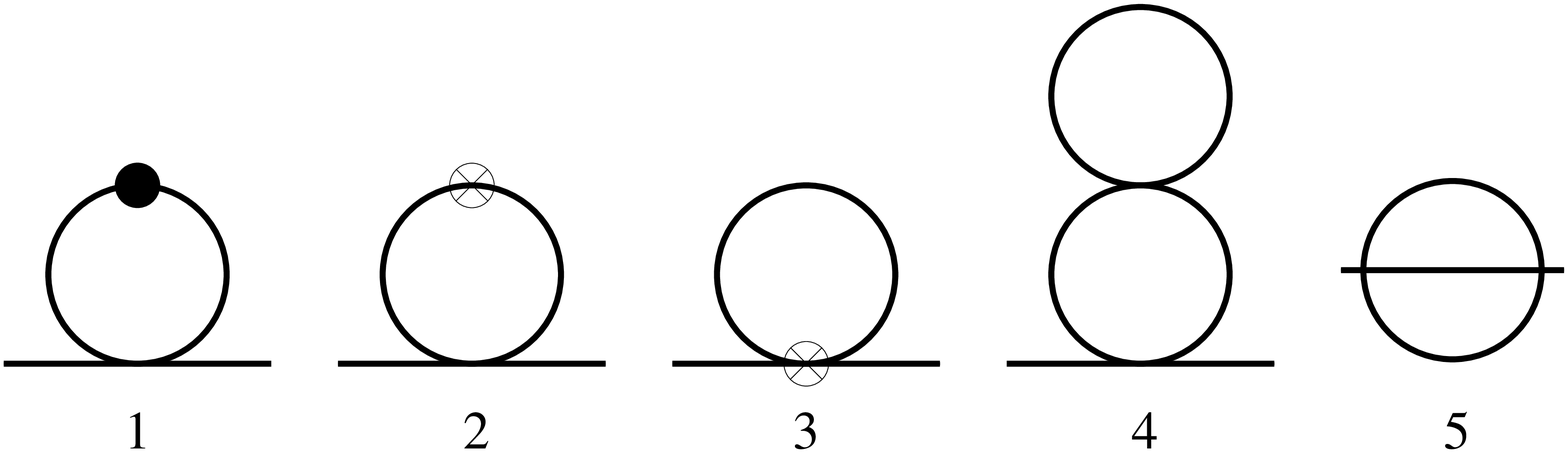}}
    \caption{Diagrams contributing at ${\mathcal O}(\delta^2)$}
%    \end{center}
\end{figure}

At ${\mathcal O}(\delta^2)$ there are five diagrams (shown in Fig.\ 2) which
provide corrections to the self-energy.  We begin by  evaluating the first
diagram, which in the high temperature limit is
\begin{equation}
    \Pi_1^{\delta^2}=\delta^2\frac{\lambda T\eta^2}{8\pi\Omega}
    \frac{1}{\sqrt{1-r^2}}-\delta^2\frac{\lambda\eta^2}{16\pi^2}
    \left[-\frac{1}{\epsilon}+\ln\left(\frac{4\pi T^2}{M^2}
    \right)-\gamma\right].
\end{equation}

The next two diagrams are constructed using mass and vertex counterterms to
give ${\mathcal O}(\delta^2)$ diagrams that eliminate the
temperature-dependent divergences arising from $\Pi_4^{\delta^2}$.
Explicitly, we obtain in the high temperature limit
\begin{eqnarray}
    \Pi_2^{\delta^2} &=& -\delta^2\frac{\lambda^2\Omega^2}{(16\pi^2)^2
    \epsilon^2}+\delta^2\frac{\lambda^2}{16\pi^2\epsilon}\left\{-\frac
    {T\Omega}{8\pi}\frac{1}{\sqrt{1-r^2}}+\frac{\Omega^2}{16\pi^2}
    \left[\ln\left(\frac{4\pi T^2}{M^2}\right)-\gamma\right]
    \right\} \nonumber\\
    & & -\delta^2\frac{\lambda^2\Omega^2}{2(16\pi^2)^2}\bigg\{
    \left[\ln\left(\frac{\Omega^2}{4\pi M^2}\right)+\gamma\right]^2
    +\frac{\pi^2}{6}\bigg\},\\
    \Pi_3^{\delta^2} &=& -\delta^2\frac{5\lambda^2\Omega^2}
    {2(16\pi^2)^2\epsilon^2}+\delta^2\frac{5\lambda^2}{32\pi^2\epsilon}
    \left\{\frac{T^2}{12}-\frac{T\Omega}{4\pi}\sqrt{1-r^2}+
    \frac{\Omega^2}{16\pi^2}\left[\ln\left(\frac{4\pi T^2}{M^2}
    \right)-\gamma-2r^2\right]\right\} \nonumber\\
    & &-\delta^2\frac{5\lambda^2\Omega^2}{(32\pi^2)^2}\bigg\{\left
    [\ln\left(\frac{\Omega^2}{4\pi M^2}\right)+\gamma-1\right]^2
    +1+\frac{\pi^2}{6}\bigg\}.
\end{eqnarray}

The momentum-independent two-loop diagram is given by
\begin{eqnarray}
    \label{pi4}
    \Pi_4^{\delta^2} &=& \delta^2\frac{\lambda^2\Omega^2}{(16\pi^2)^2}
    \frac{1}{\epsilon^2}-\delta^2\frac{\lambda^2}{16\pi^2}\frac{1}
    {\epsilon}\left\{\frac{T^2}{12}-\frac{T\Omega}{4\pi}
    \sqrt{1-r^2}-\frac{T\Omega}{8\pi}\frac{1}{\sqrt{1-r^2}}+\frac
    {\Omega^2}{8\pi^2}\left[\ln\left(\frac{4\pi T^2}{M^2}\right)
    -\gamma-r^2\right]\right\} \nonumber\\
    & & -\delta^2\frac{\lambda^2T^3}{96\pi\Omega}\frac{1}{\sqrt{1-r^2}}
    +\delta^2\frac{\lambda^2T^2}{32\pi^2}+\delta^2\frac{\lambda^2}
    {(8\pi)^2}\left\{\frac{T^3}{3}-\frac{T\Omega}{\pi}
    \sqrt{1-r^2}-\frac{T\Omega}{2\pi}\frac{1}{\sqrt{1-r^2}}+
    \frac{\Omega^2}{4\pi^2}\left[\ln\left(\frac{4\pi T^2}{M^2}
    \right)-\gamma-2r^2\right]\right\} \nonumber\\
    & & \times\left[\ln\left(\frac{4\pi T^2}{M^2}\right)-\gamma
    \right]+\delta^2\frac{\lambda^2T\Omega}{64\pi^3}\frac{r^2}{\sqrt
    {1-r^2}}+\delta^2\frac{\lambda^2\Omega^2}{(16\pi^2)^2}\left[
    \ln^2\left(\frac{\Omega^2}{4\pi M^2}\right)+(2\gamma-1)\ln
    \left(\frac{\Omega^2}{4\pi M^2}\right)+2.4\right]
\end{eqnarray}

Finally, we have the momentum-dependent setting sun diagram, which we evaluate
with the Euclidean self-energy on shell (the details of the calculation are
provided in Appendix A):
\begin{eqnarray}
    \Pi_5^{\delta^2} &=& \delta^2\frac{3\lambda^2\Omega^2}{(32\pi^2)
    ^2}\frac{1}{\epsilon^2}+\delta^2\frac{3\lambda^2\Omega^2}{(32\pi^2)^2}
    \frac{1}{\epsilon}+\delta^2\frac{\lambda^2p^2}{(32\pi^2)^2}\frac{1}
    {2\epsilon}-\delta^2\frac{3\lambda^2}{16\pi^2\epsilon}\left\{\frac
    {T^2}{24}-\frac{T\Omega}{8\pi}\sqrt{1-r^2}+\frac{\Omega^2}{32\pi^2}
    \left[\ln\left(\frac{4\pi T^2}{M^2}\right)-\gamma-2r^2\right]
    \right\} \nonumber\\
    & & +\delta^2\frac{3\lambda^2\Omega^2}{2(16\pi^2)^2}\bigg[\ln^2\left(
    \frac{\Omega^2}{4\pi M^2}\right)+(2\gamma-{\textstyle\frac{17}{6}}
    )\ln\left(\frac{\Omega^2}{4\pi M^2}\right)+1.9785\bigg]
    -\delta^2\frac{3\lambda^2}{16\pi^2}\left[
    -\ln\left(\frac{\Omega^2}{4\pi M^2}\right)+2-\gamma\right]
    \nonumber\\
    & & \times\left\{\frac{T^2}{24}-\frac{T\Omega}{8\pi}\sqrt{1-r^2}
    -\frac{\Omega^2}{16\pi^2}\left[\ln\left(\frac{\Omega}{4\pi T}\right)
    +\gamma-{\textstyle\frac{1}{2}}+r^2\right]\right\} \nonumber\\
    & & +\delta^2\frac{\lambda^2T^2}{128\pi^2}\left[\ln\left(\frac
    {\Omega^2}{T^2}\right)+5.0669+{\textstyle\frac{3}{2}}\ln(1-r^2)
    -\frac{2r^2}{\pi^2}(2\ln2-1)-\frac{1}{\pi^2}\ln^2\left(\frac{1+r}{1-r}
    \right)\ln 2\right].
\end{eqnarray}
The divergent parts of these diagrams can all be eliminated by a suitable
choice of the counterterm in Eq.\ (\ref{SCd}). As mentioned above, we use the
minimal subtraction prescription.
\section{Results}
Having obtained an expression for the renormalized thermal mass, we can set
$\delta=1$ and obtain numerical results for this mass, $m^2_{T,\mu}(\bar
{\eta})$, where $\bar{\eta}$ is determined via the PMS condition:
\begin{equation}
    \left.\frac{\partial m^2_{T,\mu}(\eta)}{\partial\eta}\right|_{\eta
    =\bar{\eta}}=0.
\end{equation}
At ${\mathcal O}(\delta)$, $\bar{\eta}$ does not depend on the
coupling and so does not generate non-perturbative information, but
non-perturbative behaviour emerges at ${\mathcal O}(\delta^2)$.
Fig.\ 3 is a typical plot, showing a clear maximum in
$m^2_{T,\mu}(\eta)$.

\begin{figure}
%    \begin{center}
    \centerline{\epsfxsize=8cm
    \epsffile{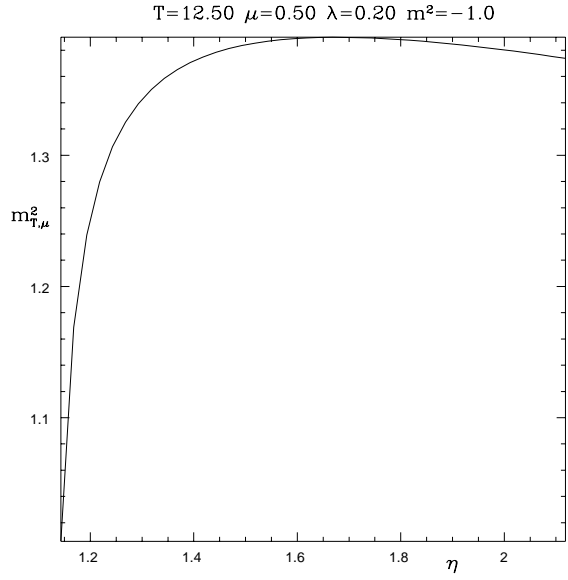}}
    \caption{Dependence of the thermal mass squared on the variational parameter $\eta$.
    All masses measured in units of $M$.}
%    \end{center}
\end{figure}

It will be of particular interest to study the $\mu$ dependence of the
critical temperature $T_c$, the signal for the phase transition being taken to
be $m^2_{T_c,\mu}=\mu^2$, which leads to an implicit equation for $T_c=T_c(\mu)$ (with
dependence on the renormalized bare mass and coupling of the theory suppressed
in the notation).  Fig.\ 4 shows a plot of the contours of $m^2_{T,\mu}$ for
$m^2=-M^2,\lambda=0.5$ in a region of the $(T,\mu)$ plane, with the thicker line
indicating the critical temperature. Below this line the LDE breaks down - the
thermal mass being a monotonically decreasing function of $\eta$ and thus
lacking extrema.

\begin{figure}
%    \begin{center}
    \centerline{\epsfxsize=8cm
    \epsffile{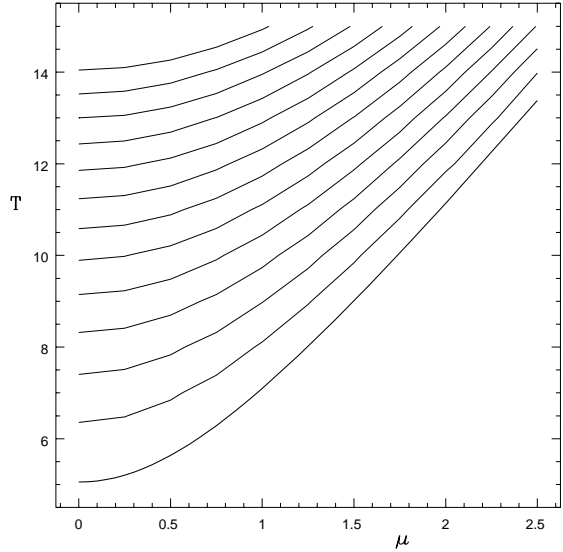}}
    \caption{Contours of the thermal mass squared in intervals of ${1\over 2}M^2$ in the $(T,\mu)$ plane for
    $\lambda=0.5$ with $m^2=-M^2$. Both axes are in units of $M$.}
%    \end{center}
\end{figure}

In Fig.\ 5 we present a comparison of $T_c(\mu)$ as calculated by the LDE
with the first-order estimate
\begin{equation}
    \label{mmu}
    m^2+\frac{\lambda T_c^2}{12}=\mu^2
\end{equation}
provided by perturbation theory in the high
temperature limit \cite{ben91}.
\begin{figure}
%    \begin{center}
    \centerline{\epsfxsize=8cm
    \epsffile{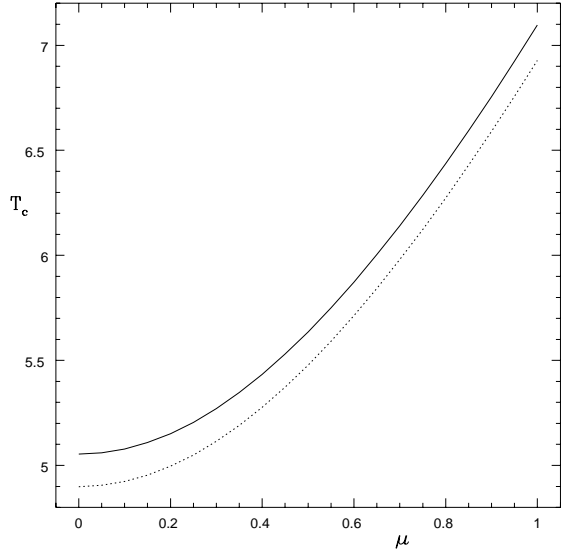}}
    \caption{Dependence of the critical temperature on $\mu$ at
    ${\mathcal O}(\delta^2)$ for $m^2=-M^2, \lambda=0.5$, with both axes in
    units of $M$. The dotted line indicates the first-order approximation given
    by Eq.\ (\ref{mmu}).}
%    \end{center}
\end{figure}

The two curves are quite similar in shape, and converge at large $T_c$.
The reason for this behaviour is that in the high $T$ limit, it so happens
that $\bar{\eta}^2\to\mu^2-m^2$, which is  equivalent to $r^2\to 1$. Examining
those contributions from $\Pi_4$ in (\ref{pi4}) we see  that if this is the
case, we require the coefficients of the $1/\surd{(1-r^2)}$ terms to disappear,
which is exactly the same condition as (\ref{mmu}).

Fig.\ 6 illustrates the high-$T$ behaviour of the second-order approximation
to the thermal mass, with $\mu/T_c$ approaching a constant as $T_c$ increases;
the constant being determined from (\ref{mmu}) to be $\surd{(\lambda/12)}$.

\begin{figure}
%    \begin{center}
    \centerline{\epsfxsize=8cm
    \epsffile{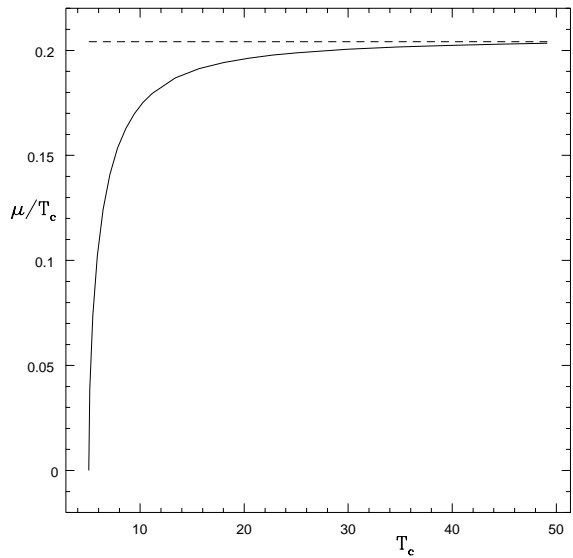}}
    \caption{The variation of $\mu/T_c$ with $T_c$ for $m^2=-M^2$,
    $\lambda=0.5$. The dashed line indicates the constant value
    $\surd{(\lambda/12)}$. $T_c$ is in units of $M$.}
%    \end{center}
\end{figure}
%%%%%%%%%%%%%%%%%%%%%%%%%%%%%%%%%%%%%%%%%%%%%%%%%%%%%%%%%%%%%%%%%%%%%%%%%%%
\section{Conclusions}
In this paper we have shown how the problem of finite chemical potential
for a charged scalar theory can be formulated in the context of the linear
delta expansion. The graphs to be evaluated are essentially those of ordinary
perturbation theory, with modified mass and coupling parameters. The chemical
potential appears  explicitly in the Lagrangian, and also in the Bose-Einstein
factors occurring in the momentum integrals. Renormalization has been
implemented on the lines of Ref.~\cite{pin99}, where the importance of
renormalizing before applying the variational aspect of the method was
emphasized. We obtain unambiguous PMS points, as illustrated in Fig.~3.

The results have been plotted in Fig.~4 as a contour plot of the thermal mass
in the $(T,\mu)$ plane, and in Fig.~5 as $T_c$ versus $\mu$ for given values
of $m^2, \lambda$. This latter curve approaches the result of resummed
perturbation theory at high temperature, but differs significantly from it
at lower temperatures. The reason for the convergence at higher temperatures
can be understood in terms of the properties of the stationary points in the
variational parameter $\eta$.

As emphasized in the introduction, the problem of a non-zero chemical potential
is not amenable to treatment by the usual Monte-Carlo lattice method. A lattice
version of the present calculation is in progress.

An extension of the present work which we intend to pursue in the near future
is to free it from the dependence on the high temperature expansion, thereby
enabling us to consider a system with non-zero chemical potential at low
temperature.
\section*{acknowledgments}
We are very grateful to Dr.~T.~S.~Evans for useful discussions and the
benefit of his expertise in thermal field theory. One of us (PP) wishes to
thank the Particle Physics and Astronomy Research Council of the UK for
financial support.
%%%%%%%%%%%%%%%%%%%%%%%%%%%%%%%%%%%%%%%%%%%%%%%%%%%%%%%%%%%%%%%%%%%%%%%%%%

\appendix
\section{}
\begin{figure}
%    \begin{center}
    \centerline{\epsfxsize=5cm
    \epsffile{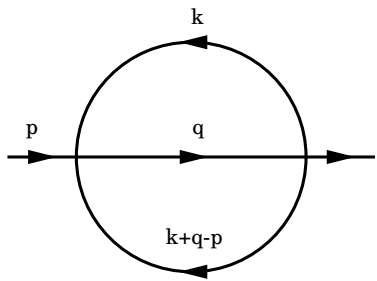}}
    \caption{The two-loop momentum-dependent diagram}
%    \end{center}
\end{figure}

In this appendix we will derive in some detail the expression for the
two-loop setting sun diagram.  We begin by mimicking the approach of
Parwani~\cite{par92} to split up the diagram into three parts; containing
zero, one and two Bose-Einstein factors respectively.  Explicitly,
\begin{equation}
    \Pi_5^{\delta^2}=-\frac{\delta^2\lambda^2}{2}M^{4\epsilon}
    T^2\sum_{l,m}\int\!\frac{d^{d-1}{\mathbf k}}{(2\pi)^{d-1}}
    \frac{d^{d-1}{\mathbf q}}{(2\pi)^{d-1}}\Delta_F(i\omega_l,{\mathbf k})
    \Delta_F(i\omega_m,{\mathbf q})\Delta_F(i\omega_{-r},-{\mathbf r}),
\end{equation}
where $d=4-2\epsilon$ is the space-time dimension and ${\mathbf r}={\mathbf p}
-{\mathbf k}-{\mathbf q}$.  If we now use the mixed representation of the
Matsubara propagator defined in (\ref{Del}), the frequency sums become trivial
and eventually we have
\begin{equation}
    \Pi_5^{\delta^2}(i\om{n},{\mathbf p})=-\frac{\delta^2\lambda^2}{2}\Big(
    G_0(i\om{n},{\mathbf p})+G_1(i\om{n},{\mathbf p})+G_2(i\om{n},
    {\mathbf p})\Big),
\end{equation}
where
\begin{eqnarray}
    G_0(i\om{n},{\mathbf p}) &=& \int\!d\,[{\mathbf k},{\mathbf q}]
    S_{\mu}(\om{k},\om{q},\om{r}),\\
    G_1(i\om{n},{\mathbf p}) &=& \int\!d\,[{\mathbf k},{\mathbf q}]\Big[
    n_k^+\Big(S_{\mu}(\om{k},\om{q},\om{r})+S_{\mu}(-\om{k},\om{q},\om{r})
    +S_{\mu}^+(\om{k},\om{q},\om{r})+S_{\mu}^-(-\om{k},\om{q},\om{r})\Big)
    \nonumber\\
    & &+n_k^-\Big(S_{\mu}(\om{k},\om{q},\om{r})+S_{\mu}(-\om{k},\om{q},
    \om{r})+S_{\mu}^+(-\om{k},\om{q},\om{r})+S_{\mu}^-(\om{k},\om{q},
    \om{r})\Big)\Big],\\
    G_2(i\om{n},{\mathbf p}) &=& \int\!d\,[{\mathbf k},{\mathbf q}]\Big[
    n_k^+n_q^+\Big(S_{\mu}^+(\om{k},\om{q},\om{r})+S_{\mu}(-\om{k},\om{q},
    \om{r})+S_{\mu}(\om{k},-\om{q},\om{r})-S_{\mu}^+(\om{k},\om{q},-\om{r})
    \Big) \nonumber\\
    & &+n_k^+n_q^-\Big(S_{\mu}(\om{k},\om{q},\om{r})+S_{\mu}^-(-\om{k},
    \om{q},\om{r})+S_{\mu}^+(\om{k},-\om{q},\om{r})-S_{\mu}(\om{k},\om{q},
    -\om{r})\Big) \nonumber\\
    & &+n_k^-n_q^+\Big(S_{\mu}(\om{k},\om{q},\om{r})+S_{\mu}^+(-\om{k},
    \om{q},\om{r})+S_{\mu}^-(\om{k},-\om{q},\om{r})-S_{\mu}(\om{k},\om{q},
    -\om{r})\Big) \nonumber\\
    & &+n_k^-n_q^-\Big(S_{\mu}^-(\om{k},\om{q},\om{r})+S_{\mu}(-\om{k},
    \om{q},\om{r})+S_{\mu}(\om{k},-\om{q},\om{r})-S_{\mu}^-(\om{k},\om{q},
    -\om{r})\Big)\Big],
\end{eqnarray}
with the definitions
\begin{eqnarray}
    S_{\mu}^{\pm}(\om{k},\om{q},\om{r}) &=& \frac{1}{\pm(i\om{n}+\mu)+
    \om{k}+\om{q}+\om{r}},\nonumber\\
    S_{\mu}(\om{k},\om{q},\om{r}) &=& S_{\mu}^+(\om{k},\om{q},\om{r})
    +S_{\mu}^-(\om{k},\om{q},\om{r}),\nonumber\\
    d\,[{\mathbf k},{\mathbf q}] &=& M^{4\epsilon}
    \frac{d^{d-1}\mathbf k}{(2\pi)^{d-1}}\frac{d^{d-1}\mathbf q}
    {(2\pi)^{d-1}}\frac{1}{8\omega_k\omega_q\omega_r}.\nonumber
\end{eqnarray}
We choose to evaluate the self-energy on shell (${\mathbf p}={\mathbf 0},
i\om{n}=\Omega-\mu$). The reason we evaluate at this point rather than at
${\mathbf p}={\mathbf 0}, i\om{n}=\Omega$ is best explained as follows\cite{TSE}.
In setting up the theory, we are dealing with an effective Hamiltonian,
$\hat{H}_{\scriptsize{\mathrm{eff}}}=\hat{H}-\mu\hat{Q}$ rather than the real
Hamiltonian, $\hat{H}$. We would naturally choose to evaluate the self-energy
at $i\om{n}=\Omega$ for the real Hamiltonian, which means evaluating at
$i\om{n}=\Omega-\mu$ in the effective theory. Following this prescription,
we find that
\begin{eqnarray}
    \Re[G_0(\Omega-\mu,{\mathbf 0})] &=& -\frac{3\Omega^2}{2(16\pi^2)^2}
    \left[\frac{1}{\epsilon^2}+\frac{3-2\gamma}{\epsilon}-\frac{2}
    {\epsilon}\ln\left(\frac{\Omega^2}{4\pi M^2}\right)\right]
    -\frac{p^2}{(32\pi^2)^2}\frac{1}{\epsilon} \nonumber\\
    & &-\frac{3\Omega^2}{(16\pi^2)^2}\left[\ln^2\left(\frac{\Omega^2}
    {4\pi M^2}\right)+(2\gamma-{\textstyle\frac{17}{6}})\ln\left(
    \frac{\Omega^2}{4\pi M^2}\right)+1.9785\right].
\end{eqnarray}

We now calculate the contribution from $G_1$. Writing
\begin{equation}
    G_1(i\om{n},{\mathbf p})=G_1^+(i\om{n},{\mathbf p})+G_1^-(i\om{n},
    {\mathbf p}),
\end{equation}
where $G_1^{\pm}$ represents that part of $G_1$ with an associated BE factor of
$n_k^{\pm}$ respectively, and decomposing this factor into a UV divergent part
and a UV finite part in the manner of the $\mu=0$ case one obtains
\begin{equation}
    \Re[G_1^\pm(\Omega-\mu,{\mathbf 0})]=F_0^\pm +F_1^\pm +F_2^\pm(\Omega^2).
\end{equation}
Here
\begin{eqnarray}
    F_0^\pm &=& \frac{3}{32\pi^2}\frac{T^2}{\pi^2}h_3(y,\pm r)\frac{1}
    {\epsilon},\\
    F_1^\pm &=& \frac{3}{32\pi^2}\frac{T^2}{\pi^2}h_3(y,\pm r)\left[\ln\left(
    \frac{4\pi M^2}{\Omega^2}\right)+2-\gamma\right],
\end{eqnarray}
and
\begin{equation}
    F_2^\pm(\Omega^2)=\frac{1}{4(2\pi)^4}\int_0^{\infty}\!dk\frac{kn_k^\pm}
    {\om{k}}\int_0^{\infty}\frac{dq}{\om{q}}\left[q\ln\left|\frac{X_+^\pm}
    {X_-^\pm}\right|-6k\right],
\end{equation}
with
\begin{equation}
    X^\pm_\pm=[\Omega^2-(\om{k}+\om{q}+\om{k\pm q})^2][\Omega^2-(-\om{k}
    +\om{q}+\om{k\pm q})^2][(\Omega\pm\om{k})^2-(\om{q}+\om{k\pm q})^2],
\end{equation}
where the subscript $\pm$ refers to the $\om{k\pm q}$ term.
Following a similar procedure to that of Ref.~\cite{par92} we find that
in the high-$T$ limit
\begin{equation}
    F_2^\pm(\Omega^2)\sim\frac{T^2}{128\pi^2}\left[\ln\left(\frac{\Omega}{T}
    \right)-0.54597\right]
\end{equation}
Thus, collecting all the terms together,
\begin{equation}
    \Re[G_1(\Omega-\mu,{\mathbf 0})]=F_0+F_1+F_2(\Omega^2),
\end{equation}
with
\begin{eqnarray}
    F_0 &=& \frac{3T^2}{16\pi^4}h_3^e(y,r)\frac{1}{\epsilon},\\
    F_1 &=& \frac{3T^2}{16\pi^4}h_3^e(y,r)\left[-\ln\left(\frac{\Omega^2}
    {4\pi M^2}\right)+2-\gamma\right],
\end{eqnarray}
and
\begin{equation}
    F_2(\Omega^2)\sim\frac{T^2}{64\pi^2}\left[\ln\left(\frac{\Omega}{T}
    \right)-0.54597\right].
\end{equation}

Finally, we consider the contribution from $G_2$, which contains a BE factor
for each loop and so is UV finite. We write
\begin{equation}
    \Re[G_2(\Omega-\mu,{\mathbf 0})]=H^{++}(\Omega^2)+H^{+-}(\Omega^2)
    +H^{-+}(\Omega^2)+H^{--}(\Omega^2)
\end{equation}
with
\begin{equation}
    H^{\pm \pm}(\Omega^2)=\frac{1}{4(2\pi)^4}\int_0^{\infty}\!dk
    \frac{kn_k^{\pm}}{\om{k}}\int_0^{\infty}\!dq\frac{qn_q^{\pm}}{\om{q}}
    \ln\left|\frac{Y_+^{\pm \pm}}{Y_-^{\pm \pm}}\right|,
\end{equation}
where
\begin{eqnarray}
    Y_{\pm}^{++} &=& [\Omega^2-(-\om{k}+\om{q}+\om{k\pm q})^2][\Omega^2-(
    \om{k}-\om{q}+\om{k\pm q})^2][(\Omega+\om{k}+\om{q})^2-\om{k\pm q}^2],
    \\
    Y_{\pm}^{+-} &=& [\Omega^2-(\om{k}+\om{q}+\om{k\pm q})^2][\Omega^2-(
    \om{k}+\om{q}-\om{k\pm q})^2][(\Omega+\om{k}-\om{q})^2-\om{k\pm q}^2],
    \\
    Y_{\pm}^{-+} &=& [\Omega^2-(\om{k}+\om{q}+\om{k\pm q})^2][\Omega^2-(
    \om{k}+\om{q}-\om{k\pm q})^2][(\Omega-\om{k}+\om{q})^2-\om{k\pm q}^2],
    \\
    Y_{\pm}^{--} &=& [\Omega^2-(-\om{k}+\om{q}+\om{k\pm q})^2][\Omega^2-(
    \om{k}-\om{q}+\om{k\pm q})^2][(\Omega-\om{k}-\om{q})^2-\om{k\pm q}^2].
\end{eqnarray}
Each of the $H^{\pm \pm}$ pieces has a logarithmic IR divergence as $\Omega
\to 0$.  To deal with this, we extract the leading order behaviour of
$\ln|Y^{\pm \pm}_+/Y^{\pm \pm}_-|$ in this limit, separating into three
terms as follows:
\begin{equation}
    \ln\left|\frac{Y^{+\pm}_+}{Y^{+\pm}_-}\right|\sim\ln\left|\frac{Y^{-\mp}_+}
    {Y^{-\mp}_-}\right|\sim\pm\ln 2\pm\ln\left|\frac{kq}{\Omega\sqrt{k^2-q^2}}
    \right|+{\textstyle \frac{3}{2}}\ln\left|\frac{k+q}{k-q}\right|.
\end{equation}
Collecting those terms involving the $\ln 2$ piece alone, which we can take
outside the integral, we calculate the coefficient of this term in the
high-temperature limit $a\equiv \beta\Omega\ll 1$, which we call $C_{\ln 2}$,
to be
\begin{equation}
    C_{\ln 2}=\frac{T^2}{(8\pi^2)^2}\ln^2\left(\frac{1+r}{1-r}\right).
\end{equation}
The remainder of $\Re[G_2(\Omega-\mu,{\mathbf 0})]$ splits into two pieces
which we define by
\begin{eqnarray}
    H^1(\Omega^2) &=& \frac{1}{4(2\pi)^4}\int_0^{\infty}\!\frac{kdk}
    {\om{k}}(n_k^+-n_k^-)\int_0^{\infty}\!\frac{qdq}{\om{q}}(n_q^+-n_q^-)
    \ln\left|\frac{kq}{\Omega\sqrt{k^2-q^2}}\right|, \\
    H^2(\Omega^2) &=& \frac{3}{8(2\pi)^4}\int_0^{\infty}\!\frac{kdk}
    {\om{k}}(n_k^++n_k^-)\int_0^{\infty}\!\frac{qdq}{\om{q}}(n_q^++n_q^-)
    \ln\left|\frac{k+q}{k-q}\right|.
\end{eqnarray}
Extracting the high-temperature limit of $H^1(\Omega^2)$ gives,
after using the symmetry of the integrand under $k\leftrightarrow q$ to
restrict the range of the $q$ integration and making the changes of variable
$\Omega\xi=\om{k},\Omega\eta=\om{q}$,
\begin{equation}
    H^1(\Omega^2)=\frac{\Omega^2}{4(2\pi)^4}\int_1^{\infty}\!d\xi
    (n_{\xi}^+-n_{\xi}^-)\int_1^{\xi}\!d\eta(n_{\eta}^+-n_{\eta}^-)
    \ln\left|\frac{(\xi^2-1)(\eta^2-1)}{\xi^2-\eta^2}\right|,
\end{equation}
where $n_{\xi}^{\pm}=1/(e^{a(\xi\pm r)}-1).$
After making the successive approximations\footnote{The second approximation in (\ref{nxi}) is
strictly valid only for $|a\xi|\ll 1$; however, its validity is assured by the
convergence of the resulting integral and has been checked numerically.}
\begin{equation}
    \label{nxi}
    n_{\xi}^+-n_{\xi}^-\approx -2ra\frac{e^{a\xi}}{(e^{a\xi}-1)^2}\approx
    -\frac{2r}{a\xi^2}
\end{equation}
in the integrand, we
have, for $a\ll 1$,
\begin{equation}
    H^1(\Omega^2)=\frac{T^2r^2}{(2\pi)^4}(\ln 2-{\textstyle\frac{1}{2}}).
\end{equation}
The calculation of $H^2(\Omega^2)$ parallels the case for $\mu=0$, leading
to
\begin{equation}
    H^2(\Omega^2)= \frac{3T^2}{64\pi^2}\left[-{\textstyle \frac{1}{2}}\ln
    \left(\frac{\Omega^2-\mu^2}{T^2}\right)-1.50699\right].
\end{equation}
%%%%%%%%%%%%%%%%%%%%%%%%%%%%%%%%%%%%%%%%%%%%%%%%%%%%%%%%%%%%%%%%%%%%%%%%%%

\end{document}